\documentclass{kapproc} 

\usepackage{psfig}
\usepackage{epsfig}

\newcommand{\msun}{M$_{\odot}$}

\newcommand{\degree}{$^{\rm o}$}

\newcommand{\ha}{H$\alpha$}

\kluwerbib

\begin{document}



\articletitle[The circumstellar structure of PMS stars]{Probing the circumstellar structure of pre-main sequence stars}


\author{Jorick S. Vink, Janet E. Drew, Tim J. Harries, and Rene D. Oudmaijer}



\begin{abstract}
We present \ha\ spectropolarimetry of a large sample of pre-main sequence (PMS) stars of low and intermediate mass, and 
argue that the technique is a powerful tool in studying the circumstellar geometry around these objects. 
For the intermediate mass (2 -- 15 \msun) Herbig Ae/Be stars we find that 16 out of 23 show a line effect, which 
immediately implies that flattening is common among these objects.
Furthermore, we find a significant difference in \ha\ spectropolarimetry behaviour between the Herbig Be and Ae groups. 
For the Herbig Be stars, the concept of an electron scattering disc is shown to be a useful concept
to explain the depolarizations seen in this spectral range. At lower masses, 
more complex \ha\ polarimetry behaviour starts to appear. The concept of a compact source of \ha\ emission that 
is formed close to the stellar surface, for instance by hot spots due to magnetospheric accretion, is 
postulated as a working hypothesis to qualitatively explain the \ha\ spectropolarimetry behaviour around Herbig Ae and 
lower mass ($M$ $<$ 2 \msun) T Tauri stars. 
The striking resemblance in spectropolarimetric behaviour 
between the T~Tauri star RY~Tau and the Herbig Ae stars suggests a common origin of the polarized line photons, and hints that 
low and higher mass pre-main sequence stars may have more in common than had hitherto been suspected.
\end{abstract}


\section{Introduction}

One of the most intriguing open issues in star formation concerns 
the formation of massive stars (e.g. Zinnecker; these proceedings).
Although there is a well-established paradigm for the formation of 
low mass T Tauri stars, namely via magnetospheric accretion, it is as yet unclear 
whether such a scenario would also apply to the more massive stars.
To be able to answer the question whether Nature allows 
the scaling-up of the formation mechanism of the Sun to the most massive stars, it first needs 
to be established whether the conditions known to prevail in the lower mass pre-main 
sequence (PMS) T Tauri stars, such as the presence of circumstellar discs and stellar magnetic 
fields, persist up to the intermediate mass (2-15 \msun) PMS 
Herbig Ae/Be stars.

The question as to whether Herbig Ae/Be stars in general are embedded 
in circumstellar discs, is still under debate. Although there are clear 
indications for flattening from millimeter imaging on larger spatial 
scales (a few hundred AU) for at least some objects (Mannings \& 
Sargent 1997), other studies, probing smaller spatial scales, 
yield results that seem contradictory. For instance, the IR interferometry 
of Millan-Gabet et al. (2001) probes scales of only a few AU, and in this regime the geometry is 
found to be rather more spherical. Nonetheless, to be able to study the 
circumstellar geometries around PMS stars at the closest spatial scales, 
one needs to resort to the tool of spectropolarimetry, as this is the only 
technique that may probe the geometry on scales of stellar 
radii (equivalent to $\sim$ 0.05 AU) compared to the $>$ 1 AU scales probed by other methods.

\section{The Tool of Linear Spectropolarimetry}

In principle, the detection of linear polarization of $\sim$ 2 \%, would teach 
us directly that a specific source is non-spherically symmetric 
on the sky. However, such a level of polarization may also be due to 
polarization by dust grains in the interstellar medium operating between 
the source and the observer. Unfortunately, properly correcting for this 
interstellar contribution has been proven to be a difficult task (e.g. McLean \& Clarke 1979). 
This is one of the prime reasons as to why spectrally-resolved polarization changes across emission 
lines are so valuable, as the interstellar polarization affects 
the continuum and the line in exactly the same way: any observed change in the polarization
across the line has to be intrinsic to the source. 

Although spectropolarimetry has widely been applied to more evolved 
early-type stars, such as classical Be stars (e.g Poeckert 1975), 
the technique has only recently been applied to pre-main sequence 
stars (Oudmaijer \& Drew 1999, Vink et al. 2002, 2003). 
For classical Be stars, the dominant effect is known to be due to 
unpolarized line emission in the presence of intrinsic continuum 
polarization (e.g. Clarke \& McLean 1974). This `depolarization' 
effect across emission lines occurs because the line photons are 
formed over a larger volume (in the circumstellar disc) than 
the continuum photons and are therefore scattered to a lesser extent 
off free electrons in the disc than are the continuum photons. 
Consequently, a drop in the polarization percentage is seen (see Fig~1a for an example).

\vspace{0.5cm}
\hbox{\psfig{figure=bd+40.ps,width=6.0cm} \hspace{0.5cm} \psfig{figure=xyper.ps,width=6.0cm}}
\hbox{\hspace{3cm} 1a) 	\hspace{6cm} 1b) \hspace{3cm}}
\small{
{\bf Figure 1:}
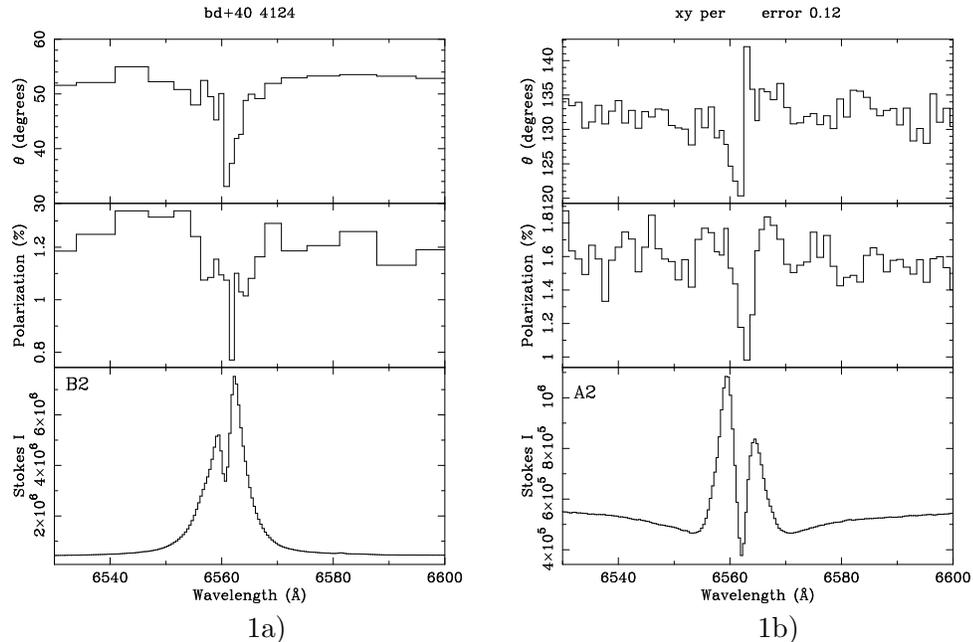
Triplots of the observed polarization spectra at \ha\ of the Herbig Be star 
BD+40~4124 (LHS) and the Herbig Ae star XY~Per (RHS). 
On both plots, the Stokes I spectrum is shown in the lowest panel,
the \%Pol is indicated in the middle panel, while the position
angle, $\theta$, is plotted in the upper panel. The data have been rebinned 
to constant errors of 0.05 \% for BD+40~4124 and 0.12 \% for XY~Per, as 
calculated from photon statistics. The data are taken from Vink et al. (2002).}
\vspace{0.5cm}

\normalsize
In certain circumstances however, it is feasible that the converse 
occurs: a proportion of the {\it line} photons originate from a compact source 
and are scattered and polarized themselves (McLean 1979; Wood et al. 1993). 
Observationally, such effects have only recently been detected in intermediate and low mass 
Herbig Ae and T~Tauri stars (Vink et al. 2002, 2003) using medium/high 
resolution ($R$ $\simeq$ 8000) spectropolarimetry. 
Here, the \ha\ line is believed to be polarized by scattering in a 
rotating non-spherically symmetric medium, most likely an accretion disc. 
Examples of both types of line effect, i.e. depolarization versus line polarization, are 
presented for respectively Herbig Be and Ae stars in Sects.~\ref{s_hbe} and~\ref{s_hae} below.

\section{The Herbig Be Stars}
\label{s_hbe}

A polarization spectrum for the Herbig Be star BD+40~4124 is shown in Fig.~1(a), and the observed 
behaviour across the \ha\ line profile in both polarization percentage (\%Pol) and position 
angle (PA) is considered to be consistent with depolarization. The reason the PA shows a change across the line
as well is attributed to the vector addition of the interstellar polarization contribution.
Note that the smooth and broad depolarization effect is represented in the $QU$ diagram of Fig.~2(a) by a more 
or less linear excursion of the line points out from the dense knot representing the continuum at $(Q,U)$ =
($-$0.3,1.25). The angle between this knot and the linear line excursion is directly related to the 
direction of the flattening of the presumed electron scattering disc around BD+40~4124.

\vspace{0.5cm}
\hbox{\psfig{figure=qu_bd+40.ps,width=6.0cm} \hspace{0.5cm} \psfig{figure=qu_xyper.ps,width=6.0cm}}
\hbox{\hspace{3cm} 1a) 	\hspace{6cm} 1b) \hspace{3cm}}
\small{
{\bf Figure 2:}
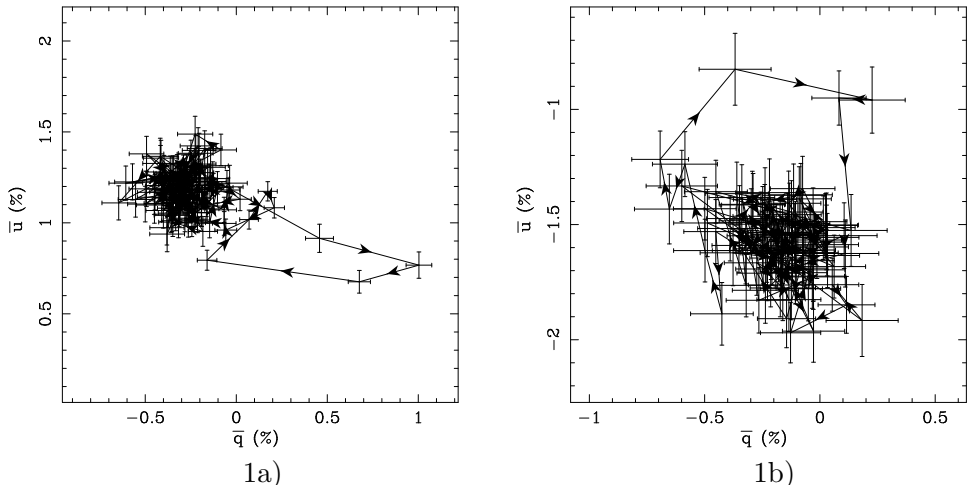
$QU$ representations of the observed polarization spectra of the same data as in Fig.~1.
The arrow denotes the sense of increasing wavelength.
The more or less {\it linear} excursion of the \ha\ line data for the Herbig 
Be star (LHS) is consistent with depolarization. The Herbig Ae data (RHS) is represented by a {\it loop} in 
the $QU$ diagram. Note that the plot axis directions $+Q$, $+U$, $-Q$, $-U$ correspond to 
sky PAs of respectively 0\degree, 45\degree, 90\degree, and 135\degree\ 
(i.e. $\Delta U/\Delta Q =$ tan $2\theta$).}
\vspace{0.5cm}

\normalsize
In the event that {\it all} Herbig Be stars are embedded in electron scattering discs, one would not 
expect a 100\% detection rate of \ha\ depolarisations, as at least some of the sources would have 
their discs too ``pole-on'' with respect to the observer to yield a \%Pol drop large enough to be detectable.
To estimate the expected fraction of depolarization detections, we turn to a comparison
with classical Be stars (Poeckert \& Marlborough 1976), objects for which the presence of a circumstellar 
disc is well-established. Applying the same detection threshold to the Poeckert \& Marlborough sample as in ours, we 
expect a detection rate of about 54\%. Returning now to our Herbig Be star 
sample, we find that 7 out of 12 (i.e. 58\%) show a detectable depolarisation (Oudmaijer \& Drew 1999, Vink et al. 2002). 
We conclude that, given both the statistics, as well as the 
smooth and broad depolarization behaviour in the Herbig Be \ha\ data, that {\it all} early Herbig Be stars are 
likely embedded in electron scattering discs. 

\section{The Herbig Ae Stars}
\label{s_hae}

When observing the later spectral type Herbig Ae stars, one may expect to see a sharp decrease in the 
frequency of line effect detections, as the circumstellar ionization as well as the amount of free electrons
that can scatter and polarize, are expected to drop among later type stars.
Another reason to expect a decrease in the frequency of line effects 
going from Herbig Be to the later Ae stars is that there appears to be a general absence of \ha\ polarization 
changes in the even later type PMS T Tauri stars (Bastien 1982; but see Sect.~\ref{s_rytau}). 

However, this turns out not to be the case at all. The number of line effects in Herbig Ae stars 
is found to be particularly high: 9 out of 11 Herbig Ae stars show a significant line effect (Vink et al. 2002).
XY~Per is included here as an example, as represented in Figs.~1(b) and~2(b): not only is there a line effect, 
it is noticeably different from the depolarization behaviour seen in the Herbig Be stars. First, the drop in the \%Pol is not as broad 
as it is for the Herbig Be stars. Second, the behaviour in PA is not smooth. Instead, a line-center flip of the PA 
is clearly noticeable in 
the upper panel of the triplot in Fig.~1(b). This PA rotation translates into a ``loop'' in the equivalent 
$QU$ diagram of Fig.~2(b).

The interpretation of these $QU$ loops in Herbig Ae stars is still a matter of 
ongoing investigation, but in the next section (Sect.~\ref{s_model}), we show that photons 
arising from a compact source of hot spots (a natural consequence of the magnetospheric accretion model) on the 
stellar surface that are subsequently scattered in a rotating circumstellar disc may 
explain the observed \ha\ spectropolarimetry data of Herbig Ae (and possibly T~Tauri 
stars; see Sect.~\ref{s_rytau}).

\section{Polarimetric Line Profiles from a Spotty Star} 
\label{s_model}

We employ a 3D Monte Carlo scattering model {\sc torus} (Harries 2000) to simulate 
a compact source of photons arising from two diametrically opposed accretion hot spots onto a  
circumstellar disc. The disc has an inner hole of 3 stellar radii, and is assumed to be flat. 
As the \ha\ emission is compact in this model, we may expect intrinsic {\it line} 
polarization to occur. Figures~3(a) and~3(b) represent the two extreme cases for an almost edge-on (LHS) and 
an almost pole-on (RHS) disc respectively. As expected for an edge-on disc, there is a significant 
level of continuum polarization of a few percent. Due to the asymmetry in the velocity field, one hot 
spot -- positioned at an angle of 65\degree\ from the equator -- illuminates the redshifted part of the 
disc that is rotating away from us, while the other (diametrically opposed) 
hot spot illuminates the blueshifted part of the disc moving towards us. Both spots illuminate the 
disc at similar angles, so that no significant PA changes occur across the line.

\vspace{0.5cm}
\hbox{\psfig{figure=edge2.ps,width=6.0cm} \hspace{0.5cm} \psfig{figure=pole2.ps,width=6.0cm}}
\hbox{\hspace{3cm} 3a) 	\hspace{6cm} 3b) \hspace{3cm}}
\small{
{\bf Figure 3:}
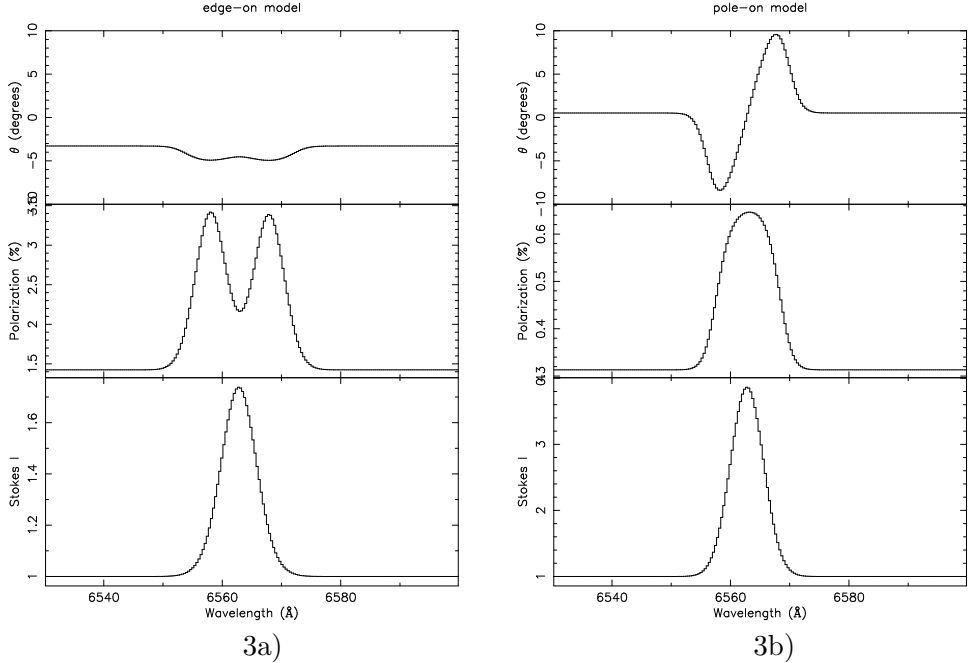
Monte Carlo models for an edge-on (LHS) and a pole-on (RHS) disc. The photon source is asymmetric applying
two diametrically opposed hot spots.} 
\vspace{0.5cm}

\normalsize
For the more pole-on case, the first difference is the lower level of continuum \%Pol, as the 
scattering is now more spherically symmetric (positive and negative $Q$ and $U$ cancel, resulting into a 
drop in the net polarization). As far as the line polarization is concerned, the asymmetry in the velocity field 
is now diminished, and this results in an effective merging of the double-peaked \%Pol profile into one single peak.
Finally, because the illumination of the pole-on disc no longer occurs under similar angles, there is a 
rotation in the PA in the upper panel of Fig.~3(b), which is indeed the same type of behaviour that is commonly 
observed in \ha\ spectropolarimetry of Herbig Ae stars (see Fig.~1b).

\section{The T~Tauri Star RY~Tau}
\label{s_rytau}

Partly because of a general absence of polarization changes across \ha\ using narrow band filters 
in T~Tauri stars (e.g. Bastien 1982), the commonly accepted view of the origin of T~Tauri 
polarization is that it is due to scattering off extended dusty envelopes.  

However, in Figure 4(a) we show the first spectropolarimetric measurements of an object classified 
as a T~Tauri star, and we notice similar complexity across \ha\ for RY~Tau as seen in 
Herbig Ae stars. Most notable is the rotation in the PA, which translates into a loop when 
plotted in $QU$ space (Fig.~4b).
This change in the polarization percentage and the PA across 
\ha\ suggests that line photons are scattered in a 
rotating disc. We are able to derive the value of the PA from the slope 
of the loop in the $(Q,U)$ diagram and find it to be 146 $\pm$ 3\degree. 
This is close to perpendicular with respect to the PA  
of the disc of 48 $\pm$ 5\degree\ as deduced from submillimeter imaging by Koerner \& Sargent (1995). 
These findings are consistent, as the PA of the imaged millimeter disc 
is expected to lie at 90\degree\ to the scattering PA deduced from the polarization data. 

We finally note that if these medium/high resolution ($R$ $\sim$ 8000) data were averaged -- as 
in the narrow \ha\ filter observations of Bastien (1982)-- they would most likely 
have produced a null result.

\vspace{0.5cm}
\hbox{\psfig{figure=rytau_binned.ps,width=6.0cm} \hspace{0.5cm} \psfig{figure=rytau_qubinned.ps,width=6.0cm}}
\hbox{\hspace{3cm} 4a) 	\hspace{6cm} 4b) \hspace{3cm}}
\small{
{\bf Figure 4:}
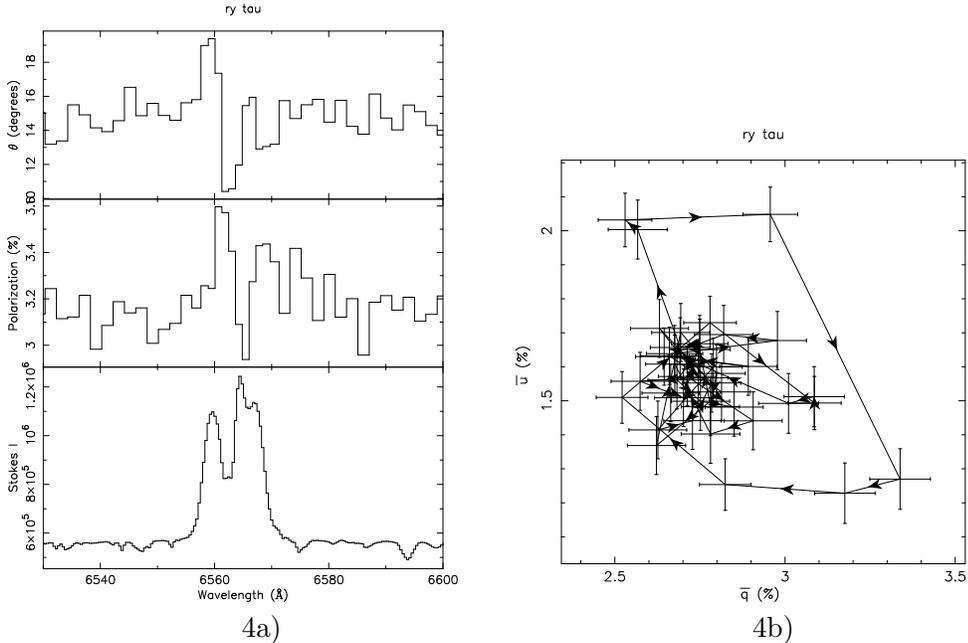
Triplot (LHS) and $QU$ diagram (RHS) of the observed polarization spectrum of the classical T~Tauri star 
RY~Tau. The data are binned to a constant error of 0.09 \%. The RY~Tau data are presented in Vink et al. (2003).
Note the flip in PA and the corresponding loop in $QU$ space.}
\vspace{0.5cm}

\section{Summary}

\normalsize
\ha\ spectropolarimetry has been shown to be a powerful tool in studying the circumstellar geometry around 
low and intermediate mass PMS stars. For the Herbig Ae/Be stars we found that 16 out of 23 show a line effect, which 
immediately implies that flattening is common among intermediate mass PMS stars.
Furthermore, we noticed a clear difference in \ha\ spectropolarimetry behaviour between the Herbig Be and Ae groups. 
For the Herbig Be stars, the concept of an electron scattering disc has been shown to be able to explain the 
depolarizations. At lower masses, more complex \ha\ polarimetry behaviour starts to appear. The concept of 
a compact source of near-stellar \ha\ emission, for instance associated with magnetospheric accretion, 
has been proposed to qualitatively explain the linear polarization changes across \ha\ seen in Herbig Ae and 
RY~Tau. 
The striking resemblance in spectropolarimetric behaviour 
between these PMS stars may suggest a common origin for the polarized line photons, and hint that 
low and higher mass pre-main sequence stars may have rather more in common than had hitherto been suspected.



%

\end{document}